# Persistent high-energy spin excitations in iron pnictide superconductors

K. J. Zhou[1*], Y. B. Huang[2,1], C. Monney[1], X. Dai[2], V. N. Strocov[1], N. L. Wang[2], Z. G. Chen[2], Chenglin Zhang[3], Pengcheng Dai[3,2], L. Patthey[1], J. van den Brink[4], H. Ding[2], and T. Schmitt[1*]

[1]*Paul Scherrer Institut, Swiss Light Source, CH-5232 Villigen PSI, Switzerland*

[2]*Beijing National Laboratory for Condensed Matter Physics, and Institute of Physics, Chinese Academy of Sciences, Beijing 100190, China*

[3]*Department of Physics and Astronomy, The University of Tennessee, Knoxville, Tennessee 37996, USA*

[4]*Institute for Theoretical Solid State Physics, IFW Dresden, 01171 Dresden, Germany*

*e-mail: kejin.zhou@psi.ch; thorsten.schmitt@psi.ch

**Motivated by the premise that superconductivity in iron-based superconductors is unconventional and mediated by spin fluctuations[1], an intense research effort has been focused on characterizing the spin excitation spectrum in the magnetically ordered parent phases of the Fe-pnictides[2-3] and – chalcogenides[4]. For these undoped materials it is well-established that the spin excitation spectrum consists of a sharp, highly dispersive magnon spanning an energy range of ~200 meV (ref. 3). The fate of these high-energy magnetic modes upon sizable doping is hitherto unresolved. Using resonant inelastic x-ray scattering we show that optimally doped superconducting $Ba_{0.6}K_{0.4}Fe_2As_2$ retains well-defined, dispersive high-energy modes of magnetic origin. These paramagnon modes are softer than, though as intense as, the magnon of undoped antiferromagnetic $BaFe_2As_2$. The persistence of spin excitations well into the superconducting phase suggests that, if spin fluctuations are responsible for superconducting pairing, they originate from a distinctly correlated spin-state. This connects Fe-pnictide superconductors to the high-Tc cuprates, for which in spite of fundamental differences in the electronic structure, similar paramagnon modes are present[5].**

One of the greatest challenges presented by the recently discovered Fe pnictide superconductors[6] is to identify the mechanism by which electrons pair when they condense into the superconducting state. Phonon-driven pairing, as for conventional superconductors, is counter-indicated by theoretical calculations[7] and the electron-phonon interaction is deemed too weak to account for the observed high critical temperatures. Attention has, therefore, turned to other types of excitations that could cause the pairing. One such candidate is the spin fluctuations[1,8] that emerge from the interactions between electron spins. For these to become operative in pairing electrons and to induce superconductivity, first the long-range magnetic ordering – the spin density wave (SDW) phase that is stable in the parent Fe-pnictides – has to melt. The subsequent crucial question is how well-defined the resulting spin-fluctuations are, which directly relates to the strength of magnetic short-range order and coherence of its magnetic excitations.

In cuprates, for which magnetic properties are governed by strong superexchange interactions between localized spin-1/2 moments in a single Cu $3d_{x^2-y^2}$ orbital, doping-induced melting of the antiferromagnetic (AF) ordered phase gives way to a highly correlated spin-liquid state. This spin-liquid carries well-defined high energy paramagnon modes, present also in optimally doped systems[5]. For the metallic Fe-pnictides, one widely held view is that the magnetic instability is due to a nesting of hole and electron Fermi-surface (FS) pockets, causing instability of the FS so that weak exchange interactions are enough to create long-range magnetic ordering[9-10]. The question arising from this is how far destabilization of the SDW ordering by doping and thereby destroying the FS nesting preserves short-range spin correlations.

In the present study with resonant inelastic x-ray scattering (RIXS) at the Fe $L_3$-edge we establish in optimally doped $Ba_{0.6}K_{0.4}Fe_2As_2$ (BKFA) the presence of well-defined and dispersive high-energy paramagnons. Their spectral intensity and width are comparable to the magnon of undoped antiferromagnetic $BaFe_2As_2$ (BFA), reaching the energy of 150 meV at the Brioullin zone (BZ) edges. These paramagnon peaks reveal the presence of a distinctly correlated spin-state in BKFA. Its existence connects the Fe-based superconductors conceptually to the high-Tc cuprates, where superconductivity also emerges in the presence of pronounced paramagnon background.

To measure spin excitations in parent BFA and superconducting BKFA Fe-pnictide samples we employ

RIXS, which has recently been established as a powerful probe of the dispersion of magnetic excitations in a wide energy-momentum window. Many studies on undoped parent cuprates have demonstrated the sensitivity of RIXS to single-magnon excitations[11-15]. Doped superconducting cuprates exhibit intense paramagnons, damped spin excitations over much of the BZ with dispersions and spectral weights closely similar to those of magnons in undoped AF ordered parent systems[5,13].

The BFA and BKFA samples used in our RIXS experiments are single crystals grown using the self-flux method[16,17]. Resistivity and bulk magnetic susceptibility measurements demonstrate the high quality of all samples (see Supplementary Information). Figures 1a and 1b display the schematics of the RIXS experimental geometry, as well as the reciprocal space which can be reached with Fe $L_3$-RIXS. A typical Fe $L_3$ edge X-ray absorption spectrum (XAS) of BFA is shown in Figure 1c, in good agreement with a previous report[18]. In Figs. 1d and 1e, a set of momentum-resolved Fe $L_3$ RIXS spectra of BFA using $\pi$ polarized incoming light at (0,0), (0.5,0) and (0.35,0.35) in the BZ are displayed. All these spectra exhibit intense Fe 3d fluorescence at around -2 eV energy transfer which has been observed in RIXS studies on other Fe pnictides[18] and chalcogenides[19]. In addition to these previous works[18,19], we reveal near the BZ edges well-defined excitations centered around 200 meV next to the quasi-elastic peak in the vicinity of zero energy.

In Figs. 2a and 2b, we show two sets of Fe $L_3$ RIXS spectra of BFA along two high symmetry directions, (0,0) - (1,0) and (0,0) - (1,1). All RIXS spectra for both BZ directions display well defined excitations within an energy range of 0-300 meV superimposed on the tail of the Fe 3d fluorescence. For high momentum transfer $q_{//}$, these excitations clearly separate from the quasi-elastic peak. Approaching the $\Gamma$ point the excitation intensity decreases and the energy position shifts towards the quasi-elastic peak. To quantitatively analyze these excitations, we subtract the fluorescence background employing the method introduced in Ref. 19 and decompose the spectral response close to the quasi-elastic peak (Supplementary Information). As demonstrated in Fig. 2c the excitation at the zone boundary peaks at around 200 meV and contains a high-energy tail. In Figs. 2d and 2e we show that the corresponding sets of excitations clearly disperse as a function of transferred momentum after subtraction of background and quasi-elastic peak. For the RIXS spectra excited with $\sigma$ polarized incoming light, the excitation intensity is slightly suppressed.

Furthermore, the spectral weight is almost quenched when the incident energy is moved away from the Fe $L_3$ resonance (Supplementary Information). These two observations are representative characteristics of single-magnon excitations as revealed with Cu $L_3$ RIXS for many cuprates[5,11-13]. Unlike the parent cuprates, which are long-range ordered AF Mott insulators, the Fe-based parent compounds are AF ordered SDW semi-metals with compensating electronlike and holelike Fermi surface pockets involving several Fe 3d orbitals. Thus, charge excitations (electron-hole pair excitations) can fall in the same energy window as spin excitations[20]. However, the electron-hole continuum is expected to be temperature independent[21]. In contrast, our RIXS measurements clearly revealed that the sharp excitations in the AF phase become much less well defined in the paramagnetic phase thereby strongly suggesting the magnetic origin of such excitations (see Supplementary Information). Comparison of our RIXS data for the parent BFA with available neutron scattering results clearly shows that the inelastic x-ray response is dominated by magnetic excitations, which is not unexpected as in direct RIXS spin-flip scattering is strong[22,23]. Fits to the inelastic response with an asymmetrical Lorentzian line shape convoluted with a Gaussian resolution function give a good description of the data[5] (see Supplementary Information). In Fig. 2f we plot the dispersion of the RIXS peak energy position as a function of momentum transfer. To exclude the effect of sample-dependent variations, we confirmed these results with independent measurements on additional samples. On top of RIXS peak positions, we overlay the dispersion curve of spin excitations extracted from INS measurements on a BFA parent sample[24] (Supplementary Information). The excellent agreement between INS and our RIXS data indicates on a simple phenomenological basis that indeed the dispersing excitations in the inelastic response are of magnetic origin. This conclusion is further supported by the fact that the line shape of the excitations can be well fitted using the formula describing the imaginary part of the dynamical spin susceptibility. Closer comparison with the INS study on parent BFA[24] demonstrates, furthermore, that our RIXS data show similar spin excitation half width at half maximum (HWHM, *i.e.*, damping) (around 100 meV) at the zone boundary (see Fig. 4b).

Having demonstrated that Fe $L_3$ RIXS allows to measure the dispersion of spin excitations in the AF ordered state, we are well prepared to further explore how spin excitations evolve in the superconducting (SC) phase. We focus now on an optimally hole-doped $Ba_{0.6}K_{0.4}Fe_2As_2$ (BKFA) superconductor ($T_c$ = 39 K), for which high energy spin excitations have not been reported so far. By performing the same measurements as for BFA, the two corresponding sets of RIXS spectra along (0,0) - (1,0) and (0,0) - (1,1) directions of BKFA are obtained and displayed in Figs. 3a and 3b, respectively. Remarkably, similar to BFA, BKFA also shows pronounced and well defined excitations persisting up to 150 meV. Since these follow the same polarization and energy dependences as parent BFA and appear as smoothly connected to the magnetic modes in BFA, we conclude that these excitations in BKFA are likewise of magnetic origin. These spin excitations show clear dispersion along the two high symmetry directions after subtraction of background and quasi-elastic peak (see Fig. 3c). For fitting of the spectral profile the same function as for parent BFA was used. Fitted peak energy, HWHM, and integrated spectral weight as a function of momentum transfer are summarized for parent BFA and superconducting BKFA in Fig. 4.

From the comparison in Fig. 4a it is noticeable that the spin excitation energies in BKFA get softened relative to the ones in BFA. Softening of spin excitations upon doping has been observed in cuprates both in RIXS[5] and INS[25] studies. Moreover, doping induces damping of the spin excitations in cuprates due to the interaction with electron-hole excitations. Interestingly, doping of the parent BFA does not create visible further damping of spin excitations (Fig. 4b). The line width of around 100 meV HWHM is likely intrinsic, since it is nearly two and half times the total instrumental resolution of our RIXS experiment (HWHM ~ 40 meV). The observed large broadening of the spin excitations already in the parent pnictide differs significantly from the situation of the parent cuprates[5,13]. In the latter case, the RIXS instrumental resolution defines the single magnon line width due to the long magnon lifetime. The larger observed magnon line width in the parent pnictide is however not unexpected since its spin excitations, despite being well defined, are essentially damped by the interaction with itinerant electrons due to its metallic nature[26]. Carrier doping into the SC state does not necessarily add further damping of spin excitations, consistent with our RIXS observation (Fig. 4b). As shown in Fig. 4c, the total spin excitation spectral weight is preserved when crossing from the AF to the SC phase. The same effect has been discovered in a RIXS study on the YBCO family, where the integrated spectral weight of spin excitations persists in parent, under- and slightly over- doped compounds[5].

Comparison between our RIXS results on pnictides and recent RIXS measurements on curpates[5] reveals thus a notable similarity in the evolution of spin excitations upon doping. Both systems show well-defined dispersive spin excitations preserved in the SC phases with the total spectral weight being largely conserved. Those observations are consistent with the presence of pronounced short-range spin correlations in the SC phases. It is remarkable that, despite pnictides and cuprates being markedly different in their FS topology, pairing symmetry, and localization of the electronic bands, they show a striking similarity in the high-energy spin excitations within the paramagnetic doped superconducting state. This implies that ground states with extended magnetic correlations persist in both of these classes of unconventional superconductors.

**Methods**

The high-quality single crystals of $BaFe_2As_2$ and $Ba_{0.6}K_{0.4}Fe_2As_2$ used in the current study were grown by the flux method as described in Refs. 16 and 17. High-resolution RIXS experiments were performed using the SAXES spectrometer at the ADRESS beam-line of the Swiss Light Source, Paul Scherrer Institut, Switzerland. The energy and momentum resolutions were 40 meV (HWHM) and 0.01Å$^{-1}$, respectively. Samples were cleaved *in-situ* and measured in a working vacuum better than $5·10^{-10}$ mbar. All samples were aligned with the surface normal (001) in the scattering plane. XAS was measured using the total electron yield method by recording the drain current from the samples. For RIXS measurements, linear polarized X-rays were used with the incident energy tuned to 708 eV at the $L_3$-edge resonance of the Fe XAS. The absolute value of the total momentum transfer, |**q**|, is constant, since the scattering angle has been kept fixed at 130°. Momentum transfer in the a-b plane is sampled through changing the grazing incident angle. This procedure is justified by the negligible influence of the out-of-plane superexchange on the in-plane spin excitation dispersion (Supplementary Information). All RIXS spectra were normalized to the integrated Fe 3d fluorescence intensity. The RIXS results have been reproduced for at least two samples of both parent and doped materials, respectively.

**Acknowledgements**

This work was performed at the ADRESS beamline of the Swiss Light Source using the SAXES instrument jointly built by Paul Scherrer Institut, Switzerland and Politecnico di Milano, Italy. We gratefully acknowledge T. Tohyama, K. Wohlfeld and M. Daghofer for fruitful discussions. We thank for financial support through the Swiss National Science Foundation and its NCCR MaNEP.

**Author contributions**

K.J.Z., H.D. and T.S. conceived the project. K.J.Z., Y.B.H., C.M., V.N.S., L.P. and T.S. carried out the experiments. N.L.W., Z.G.C., C.L.Z., and P.C.D. fabricated samples. Y.B.H. and Z.G.C. conducted sample characterization. K.J.Z. and T.S. analysed experimental data. K.J.Z., J.v.d.B., H.D. and T.S. wrote the manuscript with particular input from X.D. and all other coauthors.

**Figure legends**

**Figure 1 | Schematics of experimental geometry and representative Fe $L_3$ XAS and RIXS spectra of BFA.**
**a,** Schematics of RIXS 130° back-scattering geometry with an included angle of 50° between the incoming and outgoing light vectors, $k_i$ and $k_f$, respectively. The sample a-b plane lies perpendicular to the scattering plane. $q_{//}$ is the projection of the momentum transfer **q** along the a-b plane. The incoming light is polarized either parallel (π) or perpendicular (σ) to the scattering plane with a grazing incident angle (θ). **b,** Schematic view of the reciprocal space which can be covered by Fe $L_3$ RIXS shaded with yellow circle. Γ, B and C in Fig. 1d and e are the reciprocal positions at which RIXS spectra were collected. Black (blue) squares represent the tetragonal (orthorhombic) Brillouin zone (BZ). All RIXS spectra use the orthorhombic BZ convention for defining relative momentum transfer values. Γ point is the structural zone center, while ΓM is the AF ordering wave vector. **c,** A representative Fe $L_3$ X-ray absorption spectrum (XAS) of BFA collected with π polarized incoming light at 15 K. The arrow denotes the fixed incident energy of the Fe $L_3$-resonance for all RIXS spectra. **d,** Three typical RIXS spectra of BFA collected at 15 K with π polarized incoming light, at $(q_x, q_y)$ = (0,0) (Γ), (0.5, 0) (B) and (0.35, 0.35) (C) using relative lattice units (r.l.u.). The relation (H, K) = $(q_x a/2\pi, q_y b/2\pi)$ is adopted where a=5.506, b=5.45 Å are the orthorhombic unit cell lattice parameters in the SDW phase. **e,** Zoom into the low energy region of Fig. 1d.

**Figure 2 | RIXS spectra of parent BFA and the spin excitation dispersion along high symmetry directions in the first Brillouin zone at 15 K. a, b,** RIXS spectra of parent BFA along (0,0) - (1,0) and (0,0) - (1,1) directions obtained at 15 K with π polarized incoming light. The numbers displayed at the right side of each spectrum denote the absolute wave vector $q_{//}$ projected along the sample a-b plane (as in **e** and **Figures 3 a, b** and **c.**). **c,** A representative spectrum at the BZ edge demonstrating the subtraction of the fluorescence background and the quasi-elastic peak in order to obtain the spin excitation component. **d, e,** Spin excitations along (0,0) - (1,0) (left) and (0,0) - (1,1) (right) directions after subtracting the background and quasi-elastic peak. The arrows mark the fitted peak centers. **f,** Spin excitation dispersions along (0,0) - (1,0) (left) and (0,0) - (1,1) (right) directions. $q_{//}$-transfer is re-scaled to r.l.u. The filled black diamonds are fitted energy positions from the data shown in **d** and **e**. Vertical bars represent errors from the fitting of the energy position of the spin excitations. The thick grey line is the calculated dispersion curve using the effective superexchange parameters from INS data[24] (see Supplementary Information).

**Figure 3 | RIXS spectra of the optimally hole-doped BKFA superconductor. a, b,** RIXS spectra of BKFA along (0,0) - (1,0) and (0,0) - (1,1) directions obtained at 15 K with π polarized incoming light. The probed $q_{//}$ range is the same as in the parent BFA sample. **c,** Spin excitations along (0,0) - (1,0) (upper panel) and (0,0) - (1,1) (lower panel) directions at 15 K after subtracting the background and quasi-elastic peak. The arrows mark the fitted peak centers. Spectra are vertically stacked according to different $q_{//}$ values.

**Figure 4 | Summary of spin excitations of BFA and BKFA in different phases. a,** Dispersions of spin excitations of BFA in AF phase, and BKFA in SC phase. For BKFA the reciprocal lattice units are calculated based on the orthorhombic notation with the same lattice parameters as for the AF ordered BFA. **b, c,** HWHM (damping) and integrated intensity of spin excitations of BFA and BKFA. The horizontal dotted line in **b** marks the HWHM of the total instrumental resolution of the RIXS experiment (40 meV).

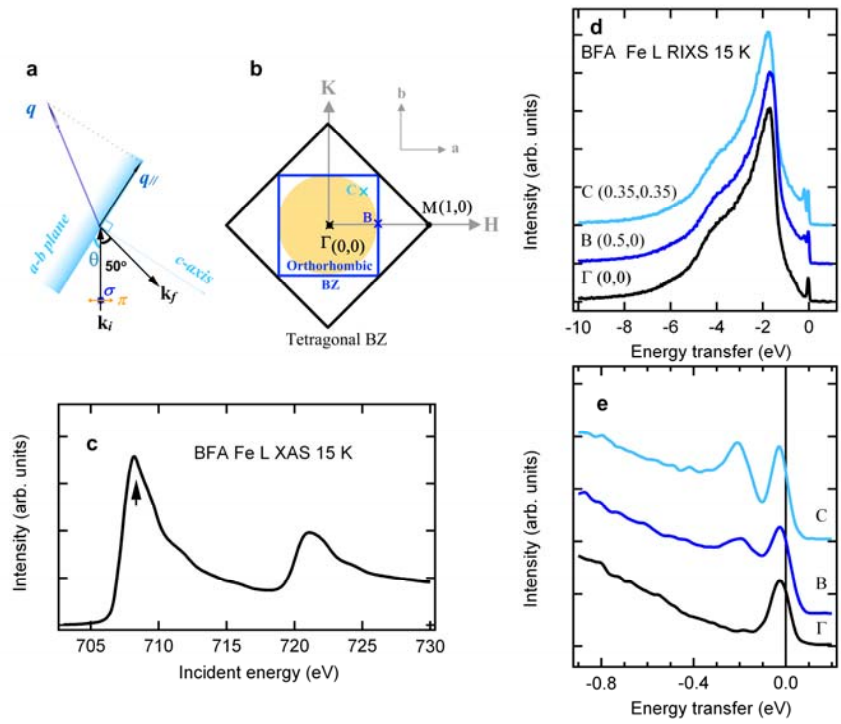

**Figure 1**

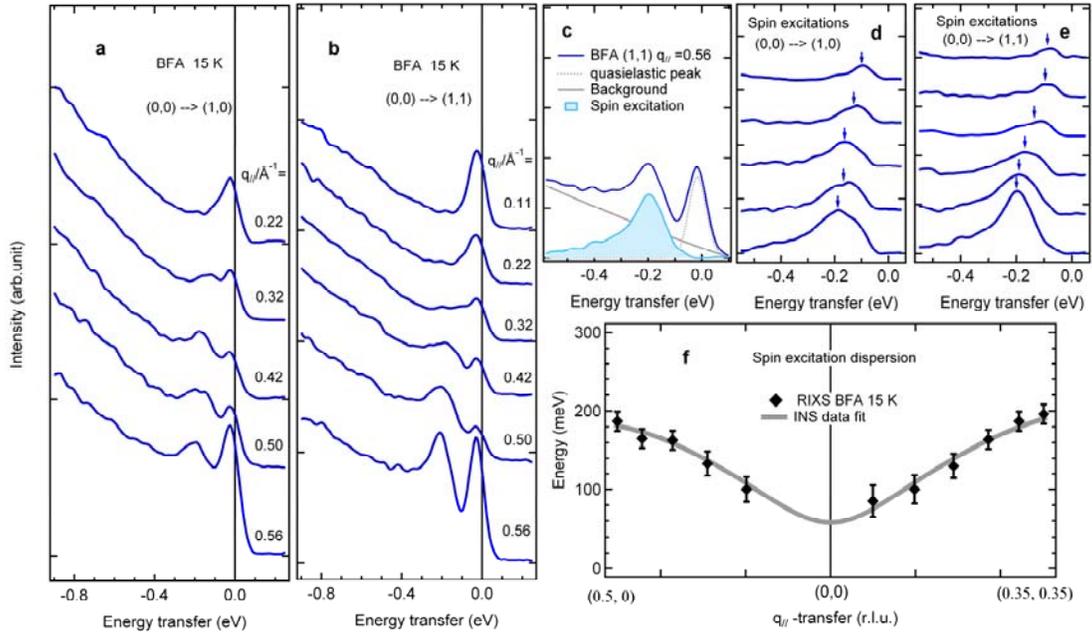

**Figure 2**

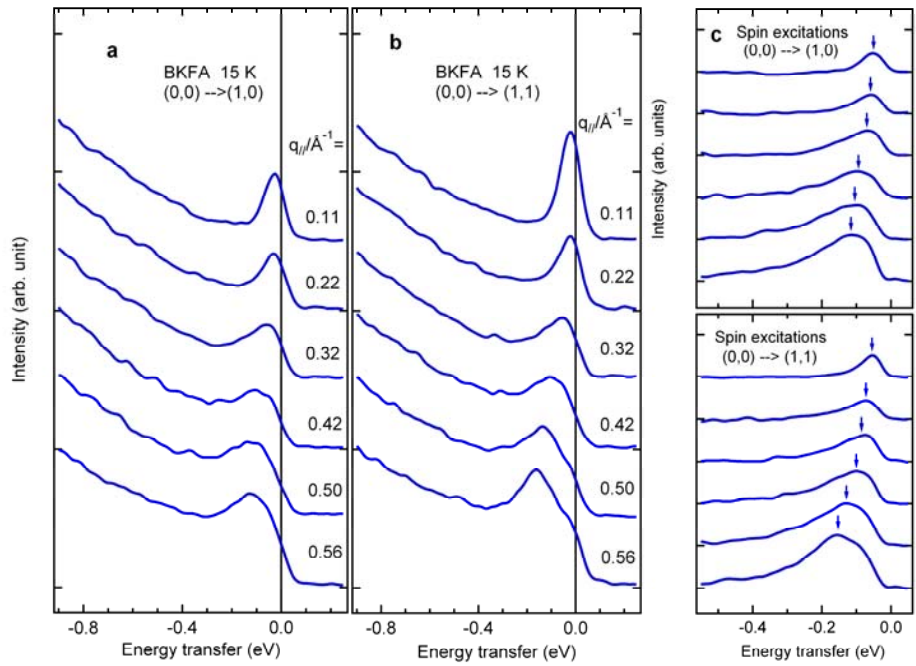

**Figure 3**

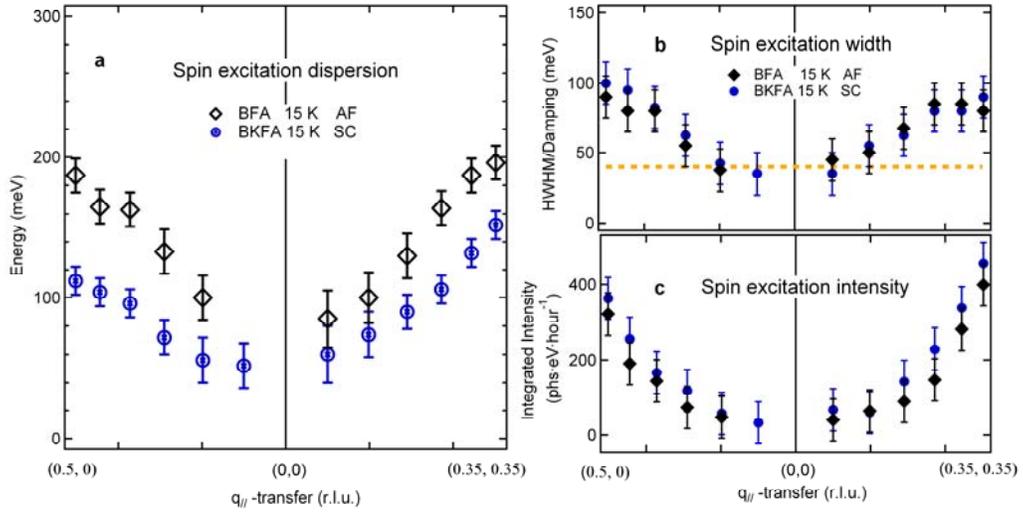

**Figure 4**

Supplementary Information for

# Persistent high-energy spin excitations in iron pnictide superconductors

K. J. Zhou[1,*], Y. B. Huang[2,1], C. Monney[1], X. Dai[2], V. N. Strocov[1], N. L. Wang[2], Z. G. Chen[2], Chenglin Zhang[3], Pengcheng Dai[3,2], L. Patthey[1], J. van den Brink[4], H. Ding[2], and T. Schmitt[1,*]

[1]*Paul Scherrer Institut, Swiss Light Source, CH-5232 Villigen PSI, Switzerland*
[2]*Beijing National Laboratory for Condensed Matter Physics, and Institute of Physics, Chinese Academy of Sciences, Beijing 100190, China*
[3]*Department of Physics and Astronomy, The University of Tennessee, Knoxville, Tennessee 37996, USA*
[4]*Institute for Theoretical Solid State Physics, IFW Dresden, 01171 Dresden, Germany*

*e-mail: kejin.zhou@psi.ch; thorsten.schmitt@psi.ch

**1. Sample growth and characterization**

The parent BaFe$_2$As$_2$ (BFA) and superconducting Ba$_{0.6}$K$_{0.4}$Fe$_2$As$_2$ (BKFA) single crystals used in this work were grown by using the self-flux method[1,2]. To characterize these samples, we performed dc resistivity and magnetic susceptibility measurements on BFA and BKFA samples, respectively. The data is shown in Fig. S1. In-plane dc resistivity was measured by a standard four-probe method and the magnetic susceptibility was measured in a Quantum Design superconducting quantum interference device vibrating-sample magnetometer system under a magnetic field of 30 Oe with H//ab plane. The cusp of the resistivity and the sharp drop of the magnetic susceptibility indicate the magnetic transition of BFA at a temperature of 134 K and a superconducting transition of BKFA at 39 K[3,1].

**2. Data analysis**

**2a. Fe 3d fluorescence background and quasi-elastic peak subtraction**

In order to fit the Fe 3d fluorescence line, we applied the procedure introduced in the analysis of RIXS spectra from a FeTe compound[4] with similar fluorescence contribution:

$$I_{fluo} = I_0[\alpha e^{-a\omega}\omega(1-g_{\Gamma_1}) + \beta e^{b\omega}g_{\Gamma_1} + \gamma e^{c\omega}g_{\Gamma_2}]$$

Above $\omega$ is the energy transfer, $\alpha e^{-a\omega}$ represents the slope of the energy region above -1 eV, $\beta e^{b\omega}$ and $\gamma e^{c\omega}$ are exponential tails, while $g_{\Gamma_{1,2}} = (e^{-(\omega-\omega_{1,2})/\Gamma_{1,2}} + 1)^{-1}$ gives rise to a smooth crossover from the quasi-linear to the exponential region with a width $\Gamma_{1,2}$ at the energy $\omega_{1,2}$. The quasi-elastic peak is fitted with two energy resolution limited Gaussian functions (Fig. S2) accounting for elastic and phonon contributions.

## 2b. Fitting the spin excitations

The spin excitation spectra in RIXS are fitted using the imaginary part of the system's spin susceptibility $\chi''(q,\omega)$ (ref.5). Since the spin excitations of the parent BFA are intrinsically broadened by the finite lifetime[6], we use for the RIXS fitting an asymmetrical Lorentzian function convoluted by the Gaussian resolution function as applied in ref. 5:

$$\chi''(q,\omega) = [\frac{\Gamma_q}{(\omega-\omega_q)^2 + \Gamma_q^2} - \frac{\Gamma_q}{(\omega+\omega_q)^2 + \Gamma_q^2}]$$

In the above formula, $\omega_q$ and $\Gamma_q$ stand for the peak energy and half width at half maximum of the spin excitation (*i.e.*, damping term), respectively. This formula is also employed for the fitting of the spin excitation data from BKFA. Examples of the fitting procedure are given in Fig. S2.

## 2c. Normalization and integration of the spin excitations

All RIXS spectra are normalized to the integrated Fe 3d fluorescence intensity in an energy transfer window of -8 eV to -1 eV. As the incident energy is always fixed at the Fe $L_3$ resonance, the Fe 3d fluorescence intensity can be used as reference for the normalization. For the integration of the spin excitation spectral weight we use an energy transfer window of -0.6 eV to 0.0 eV for all RIXS spectra. Below -0.6 eV the spectral weight from the spin excitations is negligible.

## 3. Incoming photon polarization and energy dependence of the spin excitations

For both, parent BFA and superconducting BKFA samples, spin excitations contribute in the RIXS spectra excited by $\pi$ and $\sigma$ polarized incident X-rays. $\pi$ polarized light gives rise to slightly higher spin excitation intensity. Furthermore, when the incoming photon energy is tuned away from the Fe $L_3$ resonance, the spin excitations are almost quenched for both, $\sigma$ and $\pi$ polarized incident light. These two characteristics are representative for single-magnon excitations as revealed in cuprates by Cu $L_3$ RIXS[5]. In Fig. S3 we show examples of polarization and energy dependent spin excitation spectra for the parent BFA sample. The same energy and polarization dependences occur also for the doped BKFA sample.

**4. Temperature dependence of the spin excitations**

In order to elucidate the evolution of the spin excitations across the antiferromagnetic (AF) - paramagnetic (PM) phase transition, we performed temperature dependent RIXS measurements by cooling the sample from 300 K to 15 K. RIXS spectra for three temperatures (300 K, 180 K and 15 K) are presented in Fig. S4. It is obvious that the well-defined peak of the spin excitation in the AF phase for 15 K is quenched and develops as a plateau-like broad excitation for 180 K and 300 K.

**5. Calculation of the spin excitations dispersion curve**

For understanding the momentum dispersion of the spin excitations in the RIXS spectra, we compare the spin excitations' dispersion curve of the INS data of a BFA parent compound[6] with our RIXS results. The spin dispersion curve from INS is reproduced using the same Heisenberg Hamiltonian consisting of effective in-plane nearest-neighbour ($J_{1a}$ and $J_{1b}$), next-nearest-neighbour ($J_2$), and out-of-plane ($J_c$) exchange interactions as in ref. 6. The dispersion relations are given by: $E(q_{//}) = \sqrt{A_{q_{//}}^2 - B_{q_{//}}^2}$, where

$$A_{q_{//}} = 2S[J_{1b}(\cos(\pi K) - 1) + J_{1a} + J_c + 2J_2 + J_s],$$

$$B_{q_{//}} = 2S[J_{1a}\cos(\pi H) + 2J_2 \cos(\pi H)\cos(\pi K) + J_c \cos(\pi L)].$$

In the above relations $J_s$ is the single ion anisotropy constant and $q_{//}$ is the reduced momentum transfer away from the AF zone center (1, 0, 1). (H, K, L) is defined as ($q_x a/2\pi$, $q_y b/2\pi$, $q_z c/2\pi$) in which a=5.506, b=5.45, c=12.97 Å are the orthorhombic unit cell lattice parameters in the SDW phase. To calculate the dispersion curve for the momentum space covered in our RIXS experiment, we use the fitted exchange values from ref. 6: $SJ_{1a}$ = 59.2 ± 2.0, $SJ_{1b}$ = -9.2 ± 1.2, $SJ_2$ = 13.6 ± 1.0, $SJ_c$ = 1.8 ± 0.3 meV and $J_s$ = 1.0. L is fixed to 1.0 as only negligible dispersion contributes at the center for other L values. We also take into account the twinning of domains in obtaining the dispersion curves along both directions. As is shown in the main manuscript, our RIXS data are in excellent agreement with the spin excitations dispersion curve obtained from INS within error bars[6].

When sampling the in-plane momentum transfer $q_{//}$ by varying the incidence angle only, the out-of plane value $q_{perp}$ is also changing. However, we verified in our analysis that the influence of $q_{perp}$ on the in-plane spin wave dispersion shows no difference at the boundaries and only a very small (< 10 meV) dispersion at the center, which is negligible compared to our energy resolution (see Fig. S5).

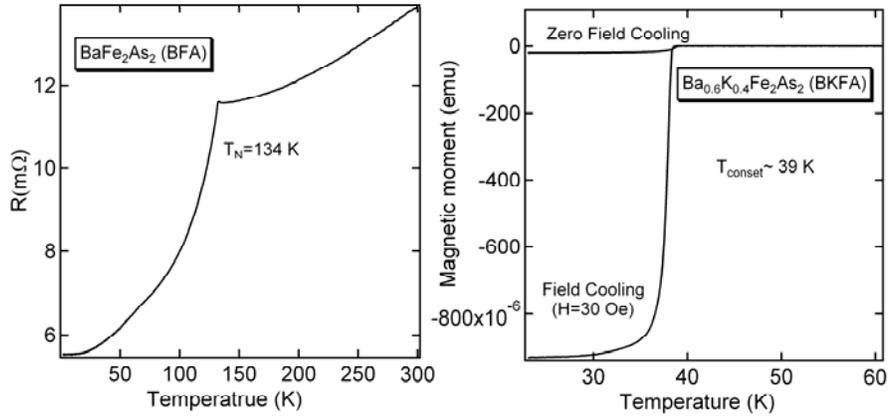

Figure S1. The dc resistivity of BFA and the magnetic susceptibility of BKFA as a function of temperature.

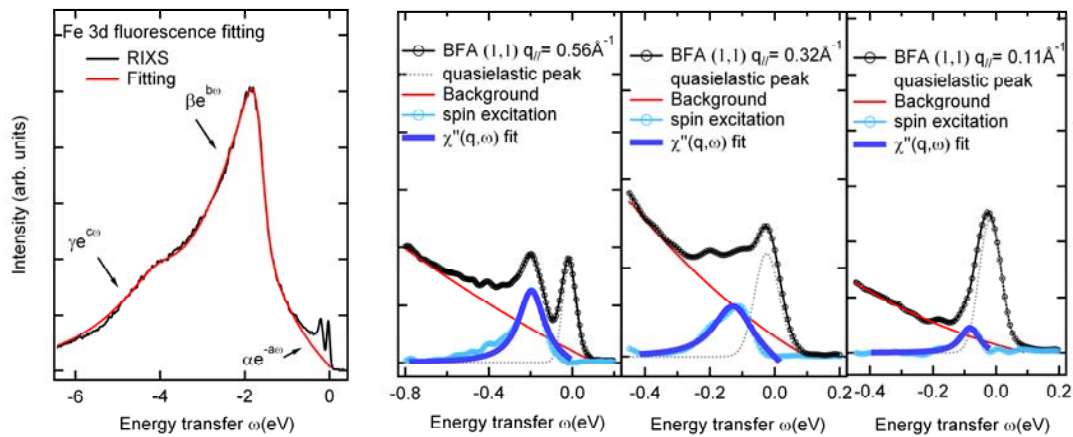

Figure S2. A representative example of a fluorescence background fit of the RIXS spectrum at $q_{//}= 0.56$ Å$^{-1}$ for parent BFA and the decomposition of the low energy response for $q_{//}$ at 0.56, 0.32 and 0.11 Å$^{-1}$ along (1,1) direction.

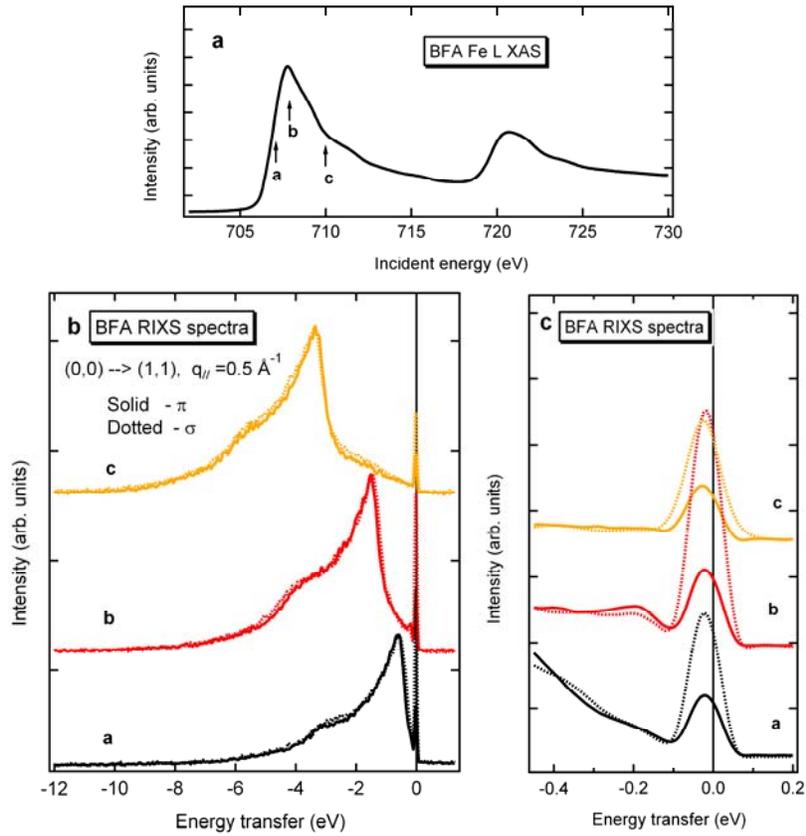

Figure S3. Incoming photon polarization and energy dependent spin excitations. **a**, A representative Fe L-edge XAS spectrum of the BFA sample collected at 15 K. **b**, Incoming photon energy and polarization dependent RIXS spectra for the BFA sample at 15 K. The three sets of spectra (vertically shifted) were recorded at different incoming photon energies as labeled in the XAS spectrum in Fig. S3a. The absolute momentum transfer is 0.5 Å$^{-1}$ along (0,0) → (1,1) direction of the orthorhombic BZ. **c**, Zoom into the spin excitations part of Fig. S3b.

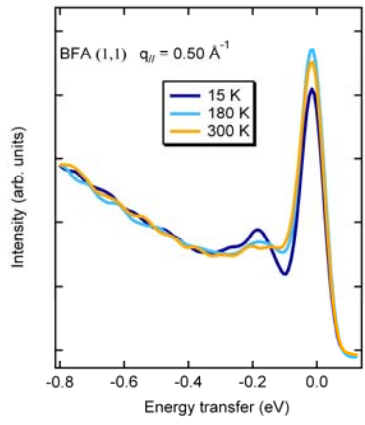

Figure S4. Temperature dependent spin excitations of the parent BFA at $q_{//}= 0.50$ Å$^{-1}$ along (1,1) direction.

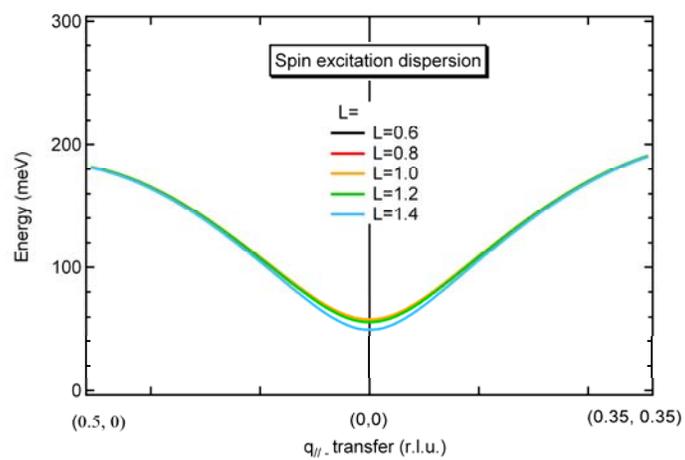

Figure S5. The influence of $q_{perp}$ on the in-plane spin excitation dispersion. L stands for the out-of plane reciprocal direction in r.l.u.